\documentclass[twocolumn]{aastex63}

\usepackage{graphicx}
\usepackage{dcolumn}
\usepackage{bm}
\usepackage{xcolor}
\usepackage{comment}
\usepackage{envmath}
\usepackage{hyperref}

\newcommand{\msun}{\ensuremath{{\rm M}_\odot}}
\newcommand{\mmax}{\ensuremath{m_\mathrm{max}}}
\newcommand{\mmaxModelB}{\ensuremath{40.8^{+11.8}_{-4.4}\,\msun}}

\newcommand{\result}[1]{\textcolor{black}{#1}}

\newcommand{\mprimary}{\ensuremath{85^{+
21}_{-14}\,\msun}}
\newcommand{\mtotal}{\ensuremath{150^{+29}_{-17} \ \msun}}
\newcommand{\msecondary}{\ensuremath{66^{+17}
_{-18} \ \msun}}

\begin{document}

\title{Minding the gap: GW190521 as a straddling binary
}

\author{Maya Fishbach}
\altaffiliation{NASA Einstein Fellow}
\email{maya.fishbach@northwestern.edu}
\affiliation{Department of Astronomy and Astrophysics, University of Chicago, Chicago, IL 60637, USA \\ Center for Interdisciplinary Exploration and Research in Astrophysics (CIERA) and Department of Physics and Astronomy, Northwestern University, 1800 Sherman Ave, Evanston, IL 60201, USA}

\author{Daniel E. Holz}
\affiliation{Enrico Fermi Institute, Department of Physics, Department of Astronomy and Astrophysics,\\and Kavli Institute for Cosmological Physics, University of Chicago, Chicago, IL 60637, USA}

\begin{abstract}
Models for black hole (BH) formation from stellar \replaced{collapse}{evolution} robustly predict the existence of a pair-instability supernova (PISN) mass gap \added{in the range $\sim50$ to $\sim 120$ solar masses}. \deleted{BHs {\em cannot}\/ be born with masses in the range $50\,M_\odot\lesssim m\lesssim 120\,M_\odot$.} This theoretical prediction is supported by the binary black holes (BBHs) of LIGO/Virgo's first two observing runs, whose component masses are well-fit by a power law with a maximum mass cutoff at $\mmax = \mmaxModelB$. Meanwhile, the BBH event GW190521 has a reported primary mass of $m_1 = \mprimary$, firmly above the inferred $\mmax$, and secondary mass $m_2 = \msecondary$. Rather than concluding that both components of GW190521 belong to a new population of mass-gap BHs, we explore the conservative scenario in which GW190521's secondary mass belongs to the previously-observed population of BHs. \added{We replace the default priors on $m_1$ and $m_2$, which assume that BH detector-frame masses are uniformly distributed}, with this population-informed prior on $m_2$, finding $m_2<\result{48}\,\msun$ at 90\% credibility. Moreover, because the total mass of the system is better constrained than the individual masses, the population prior on $m_2$ automatically increases the inferred $m_1$ to sit \emph{above} the gap (\result{39\%} for $m_1 > 120 \ \msun$, or \result{25\%} probability for $m_1 > 130 \ \msun$).
As long as the prior odds for a double-mass-gap BBH are smaller than $\sim \result{1:15}$, it is more likely that GW190521 straddles the pair-instability gap.
We argue that GW190521 may be the first example of a straddling binary black hole, composed of a conventional stellar mass BH and a BH from the ``far side'' of the PISN mass gap.
\end{abstract}

\section{Introduction} \label{sec:intro}
GW190521 is one of the most surprising and exciting systems detected thus far by the LIGO~\citep{2015CQGra..32g4001L} and Virgo~\citep{2015CQGra..32b4001A} gravitational-wave detector network. This system was detected at high confidence, with a false alarm rate of $< 1/4900$ years~\citep{GW190521}. Parameter estimation constrains the total mass of the system to be \mtotal, with a primary mass of \mprimary{} and a secondary mass of \msecondary{}~\citep{GW190521, GW190521:implications}.

Meanwhile, stellar physics predicts the existence of a BH mass gap, with no BHs in the mass range $50\,\msun \lesssim m \lesssim 120\,\msun$ due to (pulsational) pair-instability supernovae~\citep{1964ApJS....9..201F,1967PhRvL..18..379B,1983A&A...119...61O, 1984ApJ...280..825B,2003ApJ...591..288H, 2007Natur.450..390W, 2014ApJ...792...44C, 2016MNRAS.457..351Y}.
\added{The precise location of the mass gap remains theoretically uncertain, and is sensitive to details of stellar and binary evolution, including} uncertainties on nuclear reaction rates~\citep{2019ApJ...887...53F,2020arXiv200606678F,2020arXiv200913526B}, super-Eddington accretion~\citep{2020ApJ...897..100V}, convection~\citep{2020MNRAS.493.4333R}, rotation~\citep{2018ApJS..237...13L,2020A&A...640L..18M}, \added{stellar mass loss~\citep{2020ApJ...890..113B}, particularly in low-metallicity~\citep{2020arXiv201011730V} and Population III environments~\citep{2020arXiv200906585F,2020arXiv200906922K}}, and the possibility of new physics~\citep{2020arXiv200707889C,2020arXiv200901213S,2020arXiv201000254Z}. \added{Nevertheless, before the observation of GW190521, typical predictions for BHs in merging binary systems placed the lower edge of the mass gap at $\lesssim 65\,\msun$ ~\citep{2016A&A...594A..97B,2017ApJ...836..244W,2017MNRAS.470.4739S,2019ApJ...882..121S,2020ApJ...888...76M}.} Above the gap, stars of sufficiently high mass avoid pair-instability, and are expected to collapse into BHs with masses above $\sim 120$--$135\,\msun$~\citep{2016A&A...594A..97B,2016A&A...588A..50M,2019ApJ...883L..27M,2020arXiv200801890T}.

On the observational side, BBH observations in the first two observing runs (O1 and O2) of LIGO/Virgo have already placed constraints on the location of the mass gap~\citep{2017ApJ...851L..25F,o2pop,2020arXiv200807014R}. If the lower-edge of the mass gap is sharp, it is observationally measured to be $\mmax = \mmaxModelB$ (90\% credibility) using the LIGO/Virgo GWTC-1 observations~\citep{o2pop,2019PhRvX...9c1040A}, or $41^{+10}_{-5}\,\msun$ when including the IAS catalogs~\citep{2020arXiv200807014R}. It is to be noted that LIGO and Virgo are also sensitive to BHs with masses above the gap \citep{2007PhRvL..99t1102B}, and are beginning to constrain the rate of such mergers \citep{Abbott_2019,PhysRevD.102.044035}. These ``far side" black holes leave an imprint on the stochastic background of unresolved binaries, would provide unique standard siren constraints on the cosmic expansion at redshift $z\sim1$, and may also be observable by {\it LISA} \citep{ezquiaga2020jumping}.

At first glance, the primary mass of GW190521 falls squarely within this mass gap, having only 0.3\% probability of being below $65\,\msun$~\citep{GW190521}. Several scenarios, including hierarchical mergers of smaller BHs in stellar clusters or AGN disks~\citep{2002MNRAS.330..232C,2006ApJ...637..937O,2012MNRAS.425..460M,2016ApJ...831..187A,2018PhRvL.120o1101R,2017ApJ...835..165B,2020arXiv200905065F,2020arXiv200908468M,2020arXiv200609744S,2020arXiv201006161A}, primordial BHs~\citep{carr2019cosmic,2020arXiv200901728D}, stellar mergers~\citep{2019MNRAS.487.2947D, 2020MNRAS.497.1043D,2020arXiv200610771K,2020arXiv201000705R}, and accretion onto a stellar mass BH in a gas-rich environment~\citep{2020arXiv200909320S,2020arXiv200909156N, 2020arXiv200911447L} may produce BHs in the mass gap; a detailed discussion of the various possibilities is found in~\citet{GW190521:implications}.
The hierarchical merger scenario is of particular interest when one notes that the merger remnant of GW170729~\citep{2019PhRvX...9c1040A} was a BH of mass $80.3^{+14.6}_{-10.2}\,\msun$, and so LIGO/Virgo have already witnessed the creation of a BH which is consistent with the reported mass of GW190521's primary. 

However, even in these scenarios, the merger rate of systems involving a BH in the mass gap is expected to be low---typically more than two orders of magnitude smaller than the merger rate between non-mass gap BHs~\citep{2019PhRvD.100d3027R,2019MNRAS.487.2947D,2019PhRvL.123r1101Y}, especially when compared to the merger rate inferred by LIGO/Virgo~\citep{2019PhRvX...9c1040A, o2pop}. The merger rate of systems involving \emph{two} mass gap BHs is expected to be even smaller. 
Recently \citet{2020arXiv200500023K} analyzed the GWTC-1 observations under a phenomenological framework tuned to globular cluster simulations, finding that, compared to first-generation BHs, the relative rate of mergers involving one second-generation BH is $\sim 2.5\times10^{-3}$, and the relative rate of mergers involving two second-generation BHs is $\sim 3.1\times10^{-6}$. 
The relative rate of second-generation mergers from higher density environments, such as AGN disks, may be larger, but AGN disks are only expected to produce a small fraction ($\lesssim 10\%$) of LIGO/Virgo BBH events~\citep{2019PhRvL.123r1101Y}.

Because of these low expected rates, following the method of~\citet{2020arXiv200500023K}, \citet{GW190521:implications} found that a hierarchical-merger origin for GW190521 is modestly disfavored by the data by factors of $\sim 1.1$--$5$, depending on the choice of gravitational waveform model used for parameter estimation. 
\citet{GW190521:implications} also noted that the possibility that GW190521 is a first-generation BBH with $m_1$ above the PISN gap. However, they concluded that including the possibility that $m_1$ is above the gap would not significantly alter their results. Even without this possibility, the analysis of \citet{GW190521:implications} finds that both components of GW190521 are likely to be first-generation BHs.

In this paper, we build on the idea that GW190521 contains at least one conventional, first-generation BH, \added{belonging to the same population of BHs observed in LIGO/Virgo's first two observing runs}. In Section~\ref{sec:results}, we reanalyze the data with the assumption that the secondary BH is a member of the BH mass distribution inferred from LIGO/Virgo's first two observing runs, characterized by a maximum mass at $\mmax = \mmaxModelB$~\citep{o2pop}. We find $m_2 < \result{48}\,\msun$ at 90\% credibility. Because the total mass of GW190521 is constrained to be $M = \mtotal$, the updated inference on $m_2$ in turn implies that $m_1$ is likely to be the first intermediate mass black hole (IMBH) detected by LIGO/Virgo; $m_1 > 100\,\msun$ at \result{81\%} credibility. Morever, $m_1$ is likely to be on the far side of the gap: $m_1 = \result{113^{+33}_{-24}}\,\msun$ (90\% credibility), with a \result{39\%} chance that $m_1 > 120\,\msun$.
Similarly, we reanalyze GW190521 with the assumption of a PISN gap of fixed width greater than $75\,\msun$, and find that, if both component BHs avoid the gap, this naturally constrains the upper edge of the gap to be above \result{$116\,\msun$} (90\% credibility). We conclude in Section~\ref{sec:conclusions}. A derivation of the population-informed prior and a calculation of the Bayes factors between the different priors considered can be found in the Appendix.

\section{Straddling the gap}
\label{sec:results}

\begin{figure}
    \centering
    \includegraphics[width=0.5\textwidth]{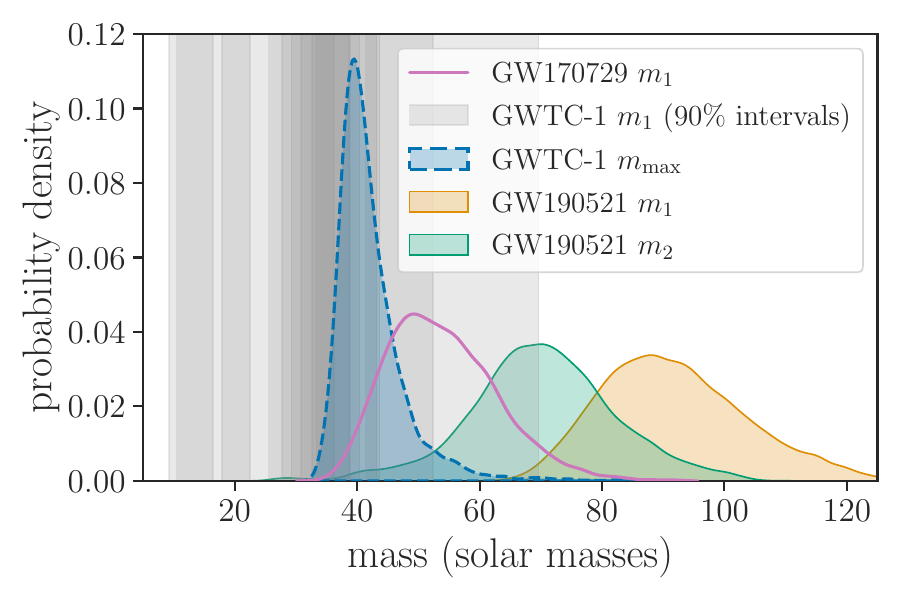}
    \caption{Posterior probability distribution for the primary and secondary mass of GW190521 (orange and green filled curves, respectively), compared to the primary masses of the ten GWTC-1 BBHs (gray bands, denoting central 90\% credible intervals), under flat, uninformative priors. The posterior on the primary mass of GW170729, the most massive BBH from GWTC-1, is additionally shown by the pink, unfilled curve. The filled, dashed blue curve shows the BH maximum mass posterior inferred from the GWTC-1 BBH events. The primary mass of GW190521 is confidently above the allowed $\mmax$ as measured from GWTC-1, suggesting it belongs to a different population. However, within measurement uncertainty, $m_2$ is consistent with being below $\mmax$, motivating our reanalysis of the GW190521 masses under the assumption that $m_2$ belongs to the previously-observed BH population.}
    \label{fig:mmax}
\end{figure}

\begin{figure*}
    \centering
    \includegraphics[width = \textwidth]{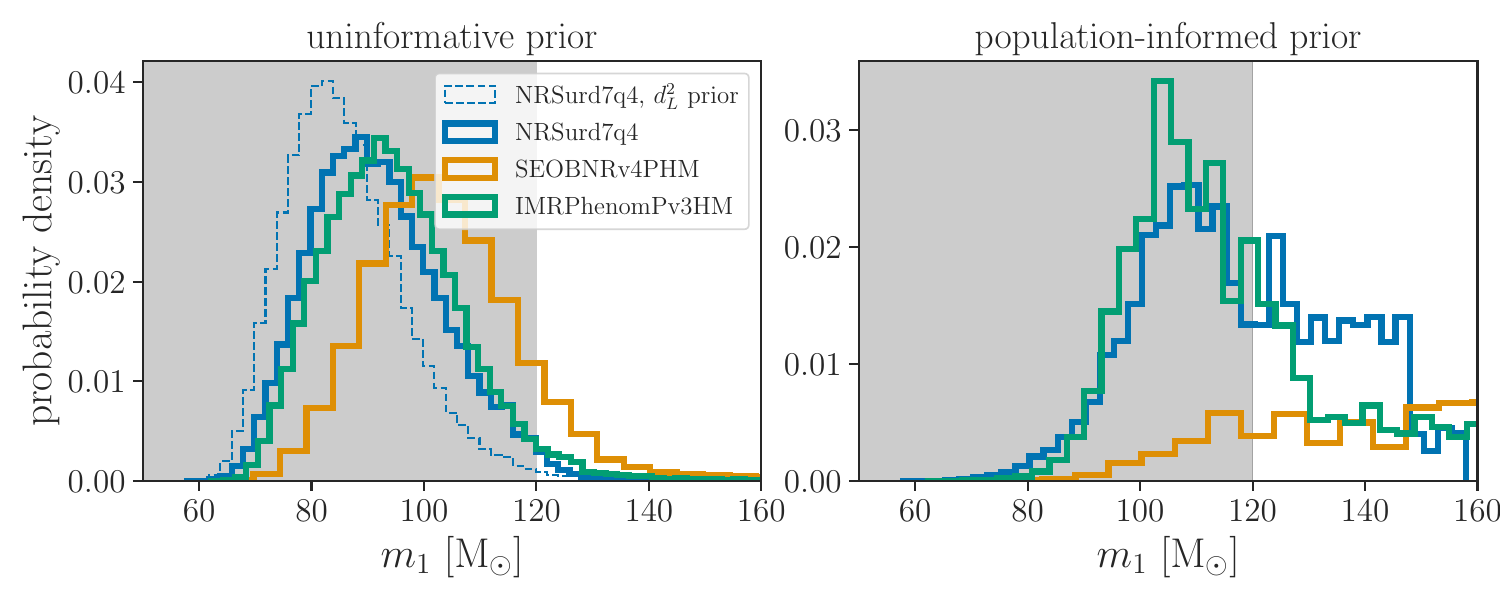}
    \caption{Posterior distribution on the source-frame primary mass using an \emph{uninformative} prior (left), compared to a population-informed prior (right) that assumes that $m_2$ belongs to the previously-observed population of BHs~\citep{o2pop}. The different colored histograms correspond to different waveform models. On the left, the solid histograms assume a flat prior on $m_1$ and $m_2$ and a flat prior on the comoving spacetime volume. The dashed blue histogram shows the posterior under the ``default" flat-in-detector-frame masses and $p(d_L) \propto d_L^2$ luminosity distance prior presented in \citet{GW190521, GW190521:implications}. On the right, we impose a prior on $m_2$ according to the component mass distribution inferred from the GWTC-1 distribution~\citep{o2pop}{, but leave the flat prior on $m_1$ (note that the new prior is on $m_2$, but we are plotting $m_1$)}.
    The shaded band denotes the region of the posterior with $m_1 < 120 \ \msun$, in which $m_1$ would be in the PISN mass gap. Under the uninformative prior, the probability that $m_1 > 120\,\msun$ is \result{1.7\%, 3.3\%, and 14\%} for the \textsc{NRSurd7q4}~\citep{2019PhRvR...1c3015V}, \textsc{IMRPhenomPv3HM}~\citep{1995PhRvL..74.3515B,2005PhRvD..71l4004B,2001PhLB..513..147D,2009PhRvD..79j4023A,2014LRR....17....2B,2020PhRvD.101b4056K}, and \textsc{SEOBNRv4PHM}~\citep{1999PhRvD..59h4006B,2000PhRvD..62f4015B,2020arXiv200409442O} waveform models respectively. Under the assumption that $m_2$ belongs to the black hole population found in GWTC-1, the probabilities for $m_1 > 120\,\msun$ increase to \result{39\%, 31\%, and 89\%} under the respective waveform models.}
    \label{fig:m1-posts}
\end{figure*}

In this section, we reanalyze the component masses of GW190521 with priors informed by the population of BBHs inferred from LIGO/Virgo's first two observing runs~\citep{o2pop}. We compare these results to the analysis of GW190521 presented in~\citet{GW190521,GW190521:implications}, which utilized broad, uninformative priors on the component masses. In particular, the priors of~\citet{GW190521,GW190521:implications} presume that the distribution of BH masses is uniform in detector-frame component masses $m_1(1+z)$ and $m_2(1+z)$, and uniform in luminosity-distance-volume ($p(d_L) \propto d_L^2$). The chosen prior ranges are such that the likelihood is contained entirely within the prior bounds.
For the remainder of this work, we employ a slightly different choice of ``uninformative" prior compared to ~\citet{GW190521:implications}. We define the uninformative prior to be flat in \emph{source}-frame masses $m_1$ and $m_2$, and uniform in comoving volume and source-frame time, $p(z) \propto dV_c/dz (1+z)^{-1}$ (see Eq.~C1 in~\citealp{o2pop}). For concreteness, we take prior bounds of $5\,\msun < m_2 < m_1 < 200\,\msun$.
Because our uninformative prior is flat over source-frame masses, the resulting posterior is proportional to the marginal likelihood.

The posterior on the component masses of GW190521 under this uninformative prior is shown by the filled orange and green curves in Fig.~\ref{fig:mmax}. We present results using the \textsc{NRSurd7q4} waveform model, the preferred waveform for this system according to~\citet{GW190521}. (A comparison of the $m_1$ posterior inferred under different waveform models can be found in Fig.~\ref{fig:m1-posts}).
Also in Fig.~\ref{fig:mmax}, we show the $m_1$ posteriors inferred for the ten BBH events of GWTC-1~\citep{2019PhRvX...9c1040A} under the same flat prior (pink unfilled curves). As discussed in the introduction, the GWTC-1 BBH events show evidence for a sharp drop in the mass spectrum at $\sim 45\,\msun$~\citep{2017ApJ...851L..25F,o2pop,2020arXiv200807014R}, consistent with expectations from PISN modeling\added{~\citep{2017ApJ...836..244W,2019ApJ...887...53F,2019ApJ...882...36M}}.
\citet{o2pop} found that the BBH mass spectrum is well-fit by a power law with a maximum mass cutoff at $\mmax = \mmaxModelB$. This $\mmax$ posterior is shown as the filled, dashed blue curve in Fig.~\ref{fig:mmax}.

It is clear from Fig.~\ref{fig:mmax} that the primary mass of GW190521 is inconsistent with a maximum mass of $\mmax = \mmaxModelB$ as inferred from the first two LIGO/Virgo runs. While the primary mass of GW190521 is greater than $64\,\msun$ at 99.95\% credibility, $\mmax$ is constrained to be less than $64\,\msun$ at 99\% credibility. On the other hand, within their measurement uncertainties, the secondary mass of GW190521 may sit below $\mmax$ as inferred from GWTC-1. While the uninformative prior finds the bulk (70\%) of the $m_2$ likelihood probability to be above $65\,\msun$, this could reasonably be explained by statistical fluctuations. 
Although $45\,\msun$ corresponds to the lower $2.6\%$ tail of the $m_2$ marginal likelihood, it is expected that one out of $\sim 40$ events will have its true mass in the 2.5\% tail.
We note that the observed primary mass of GW170729, the most massive event of GWTC-1 (see the rightmost pink curve in Fig.~\ref{fig:mmax}), also appeared larger than $\mmax$, although its true mass was likely $\lesssim 45\,\msun$~\citep{2020ApJ...891L..31F}.

We therefore consider the scenario in which the secondary mass of GW190521 belongs to the previously-observed population of BHs (implying that $m_2 < \mmax$). We refer to this as the population-informed prior. While we assume that the secondary mass of GW190521 is drawn from the BBH population inferred from GWTC-1, it is clear from Fig.~\ref{fig:mmax} that $m_1$ must belong to a different population. We therefore maintain the flat prior on $m_1$. A mathematical description of the prior can be found in Appendix~\ref{sec:methods}.

\begin{figure}
    \centering
    \includegraphics[width = 0.5\textwidth]{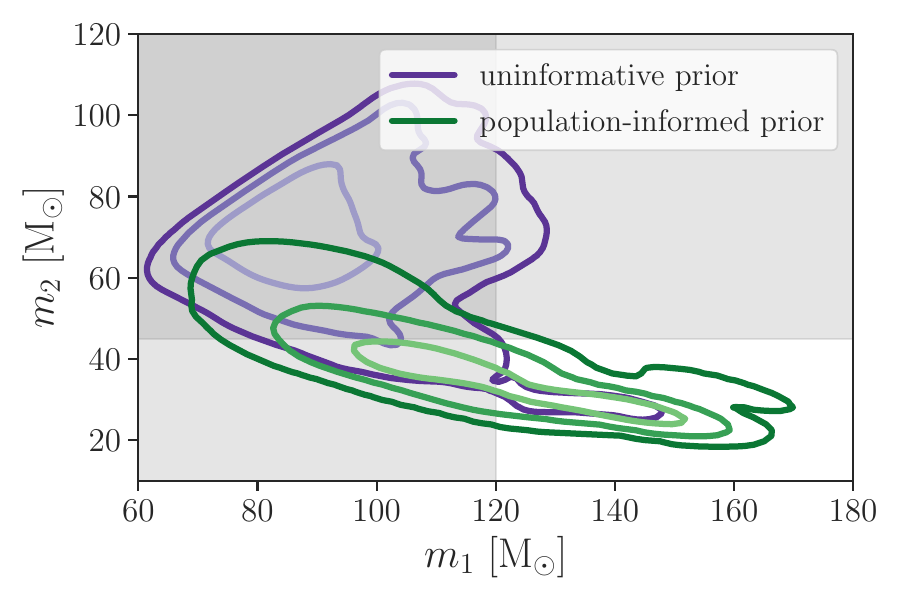}
    \caption{Two-dimensional version of Fig.~\ref{fig:m1-posts}, showing results from the \textsc{NRSurd7q4} waveform. Contours show 50\%, 90\%, and 99\% credible regions. The shaded bands show $m_2 > 45\,\msun$ (excluded if we believe that $m_2$ is a conventional BH) and $m_1 < 120 \ \msun$ (excluded if we believe that $m_1$ is a conventional BH). The unshaded region corresponds to the scenario in which the components straddle the gap.
    }
    \label{fig:m1m2joint}
\end{figure}

Under the population-informed prior, we unsurprisingly infer that $m_2<45\,\msun$ at \result{85\%} credibility.
Moreover, because the total mass of GW190521 is constrained to be $\mtotal$, the updated $m_2$ posterior affects the joint posterior on $m_1$ and $m_2$; see Fig.~\ref{fig:m1m2joint}. The implied marginal posterior on $m_1$, under the informed prior on $m_2$,
is shown in the right panel of Fig.~\ref{fig:m1-posts}.
We find that applying a population-informed prior on $m_2$ results not only in $m_2$ dropping out of the mass gap to lower values, but also results in the $m_1$ posterior increasing to \result{{$113^{+33}_{-24}$}\,\msun} and potentially crossing the upper edge of the PISN mass gap ($m_1 > 120\,\msun$ with \result{39\%} credibility, or $m_1 > 130\,\msun$ with \result{25\%} credibility). Applying the population prior on $m_2$ thus results in significant support for the two black hole masses straddling the PISN gap, with one below and one above. In Appendix~\ref{sec:Bayesfactors}, we find that the likelihood ratio between a flat prior on $(m_1, m_2)$ and a population-informed prior on $m_2$ coupled with a flat prior on $m_1 > 120\,\msun$ is of order unity, suggesting that independently of the prior odds, the data is consistent with both interpretations.

An alternative approach is to take a theoretically-motivated prior rather than a prior determined by previous observations.
The width of the gap may face fewer theoretical uncertainties than its edges;
\citet{2020arXiv200606678F} predict a width of $83^{+5}_{-8} \ \msun$. Because there is significantly more likelihood support for $m_1$ to be above $\sim100\,\msun$ than below $75\,\msun$, and the opposite holds for $m_2$, a gap width of $>75\,\msun$ naturally forces $m_1$ to be above the gap and $m_2$ to be below it.
We therefore consider a uniform prior on $m_1$ and $m_2$ with $m_1-m_2>75\,\msun$, finding $m_1>\result{116\,\msun}$ (90\% posterior probability) and $m_2<\result{41\,\msun}$ (90\% probability).
Assuming a theoretical prior on the gap width leads us to infer $m_1$ and $m_2$ values that are consistent with predictions for the gap edges.
As was the case in Fig.~\ref{fig:m1m2joint}, high values of $m_1$, which push the BH up and out of the mass gap, are accompanied by lower values of $m_2$, which cause it to drop out of the mass gap.

\section{Discussion}
\label{sec:discussion}
As discussed in the Appendix, the arguments laid out in this work are not rooted in a full population analysis, in contrast to the analysis of~\citet{2020arXiv200500023K} and \citet{GW190521:implications}.
A careful population analysis would analyze all BBH events observed thus far simultaneously, including mass, spin and redshift information, and explicitly model the distribution of mergers above the gap, taking into account upper limits on the rate of such mergers from the first two observing runs~\citep{Abbott_2019}.
Since we currently lack an observed population of BH events above the gap or theoretical guidance for the shape of the mass distribution at such high masses, we take a simple approach that applies a one-dimensional population prior to $m_2$ alone.
\added{This population prior shifts the inferred mass ratio of GW190521 to $q < 0.45$ at 90\% credibility, compared to $q > 0.45$ at 97\% credibility using flat priors on source-frame masses.}
In our population analysis, however, we do not take into account prior knowledge regarding the mass ratio distribution,
implicitly assuming that a mass ratio $q < 0.45$ is as likely as $q=1$ in the underlying population.
While this is contrary to expectations from O1 and O2, which favored equal-mass binaries~\citep{o2pop,2019MNRAS.484.4216R,2020ApJ...891L..27F,2020arXiv200807014R}, we already know from the detection of GW190412 that asymmetric systems are not uncommon, with $10\%$ of BBH systems likely having mass ratios more extreme than $q < 0.4$~\citep{GW190412}. We thus expect that under a full population analysis, the implied mass ratio of GW190521 under our proposed scenario ($q\sim 0.4$) would be reasonable. Even in the event that the underlying BBH population prefers near-unity mass ratios, and there exists a population of BHs above the gap, LIGO/Virgo at current sensitivities may be less likely to detect a $120$-$120\,\msun$ merger at cosmological distances than a system with a lower total mass similar to GW190521, as the more massive merger will merge at low frequencies out of the detectors' sensitive band~\citep{Abbott_2019,PhysRevD.102.044035}.

A future population analysis should also account for BH spins and eccentricity. \added{Hierarchical mergers leave a distinct signature on the value of the dimensionless spin magnitude $0 < \chi < 1$ of the final BH, resulting in BHs with $\chi \sim 0.7$.} More generally, dynamically-assembled binaries are expected to have isotropically-distributed spin tilts and include a fraction ($\lesssim 10\%$) of systems with measurable eccentricity. The population distributions of spin~\citep{Rodriguez_2016,2017PhRvD..96b3012T,2017ApJ...840L..24F,2017PhRvD..95l4046G,Farr:2017uvj,2017CQGra..34cLT01V,2017MNRAS.471.2801S,2018ApJ...854L...9F} and eccentricity~\citep{2018PhRvD..97j3014S,2019ApJ...871...91Z} can be used to distinguish between formation channels. In this analysis, we do not consider spin or eccentricity information. GW190521 displays mild hints of spin precession~\citep{GW190521,GW190521:implications} and/or eccentricity~\citep{2020arXiv200904771R,2020arXiv200905461G}, which may make it more consistent with a dynamical- and possibly hierarchical-merger origin. However, the preference for spin precession as measured by the $\chi_p$ parameter\added{~\citep{2015PhRvD..91b4043S}, defined as a combination of in-plane spin components,} is inconclusive~\citep{GW190521:implications}. The updated mass priors considered in this work do not significantly increase or decrease the mild preference seen for precession under the uninformative prior (for precession, $\chi_p > 0$, where $\leq 0 \chi_p \leq 1$ by definition). The uninformative prior finds $\chi_p > 0.5$ at 79\% credibility; our population prior on $m_2$, which retains an uninformative spin prior, finds $\chi_p > 0.5$ at 77\% credibility. Meanwhile, the degree of eccentricity in GW190521 is degenerate with the source-frame masses, and may further increase the probability that GW190521 contains a BH on the far side of the mass gap~\citep{2020arXiv200904771R,2020arXiv200905461G}.

\section{Conclusions}
\label{sec:conclusions}

GW190521 is one of the most surprising and important BBH merger detections to date. When analyzed with uninformative priors, both component masses of GW190521 fall within the PISN mass gap. This \replaced{implies either a breakdown of our basic understanding of stellar explosions}{challenges our understanding of stellar evolution~\citep{2020arXiv200606678F,2020arXiv200906585F,2020arXiv200901213S}}, or implies the existence of novel processes such as hierarchical mergers of smaller black holes or stellar mergers~\citep{2020arXiv200905065F,2020arXiv201000705R}. 

In this paper we have analyzed GW190521 with population-informed priors, under the assumption that, given previous BBH observations and theoretical guidance, a merger with at most one mass-gap BH is \emph{a priori} more likely than a double mass-gap merger. We have used the existing population of BBH detections to set a mass prior on the secondary BH (which has non-zero support below the gap). With this prior applied only to the secondary, we naturally find that the primary black hole has significant support {\em above}\/ the gap, making GW190521 the first observed merger between a stellar-mass BH and an IMBH. 
We also analyze GW190521 with an astrophysically-informed prior \added{that assumes that} there exists a gap of width $>75\,\msun$. In this case, we again find that GW190521 consists of a straddling binary, with component masses on either side of the gap.
Such a straddling binary fits with stellar theory, and does not necessarily require more speculative or unusual formation channels. Of course, the straddling configuration cannot exclude such alternative formation channels, as scenarios that populate the mass gap, including hierarchical mergers, may also produce BHs above the gap.

GW190521 has demonstrated that there must either be a population of BBH systems with both components within the PISN gap, or a population of component BHs with masses above the gap. 
Taking the conservative assumption that at least one of the components of GW190521 belongs to the already-observed population of BBH systems, we find that GW190521 is more likely to be a straddling binary.
Future BBH observations will help resolve this question. While the measurement uncertainty on individual events is large and thus prior-dependent, a population of BBH systems will reveal the shape of the BBH mass distribution, allowing us to firmly measure the rate of double-mass-gap binaries compared to the rate of mergers with components above the gap.

\acknowledgments
We thank Christopher Berry, Juan Calderon Bustillo, Michael Coughlin, Reed Essick, Vicky Kalogera, Ajit Mehta, Javier Roulet and members of the LIGO/Virgo collaboration for useful discussions.
MF was supported by the NSF Graduate Research Fellowship Program under grant DGE-1746045, and by NASA through the NASA Hubble Fellowship grant HST-HF2-51455.001-A awarded by the Space Telescope Science Institute. MF and DEH were supported by NSF grants PHY-1708081 and PHY-2011997, the Kavli Institute for Cosmological Physics at the University of Chicago and an endowment from the Kavli Foundation.
DEH gratefully acknowledges a Marion and Stuart Rice Award.
This research has made use of data, software and/or web tools obtained from the Gravitational Wave Open Science Center (https://www.gw-openscience.org), a service of LIGO Laboratory, the LIGO Scientific Collaboration and the Virgo Collaboration. LIGO is funded by the U.S. National Science Foundation. Virgo is funded by the French Centre National de Recherche Scientifique (CNRS), the Italian Istituto Nazionale della Fisica Nucleare (INFN) and the Dutch Nikhef, with contributions by Polish and Hungarian institutes.

\bibliographystyle{aasjournal}
\bibliography{references}

\appendix
\section{Population-informed prior}
\label{sec:methods}
This section explains the population-informed prior that we apply to the masses of GW190521 in Section~\ref{sec:results}. 
This prior is motivated by the population of BBHs observed in LIGO/Virgo's first two observing runs.
As recently discussed in~\citet{2020ApJ...891L..31F, 2019arXiv191209708G, 2020ApJ...895..128M}, combining multiple events in a population analysis allows us to update our inference on the parameters of the individual events. By learning the population distribution, a hierarchical Bayesian framework updates the prior distribution of the individual-event parameters, allowing us to self-consistently infer the shape of the population distribution (i.e., the mass and spin distributions) jointly with the parameters of individual events~\citep{2010PhRvD..81h4029M}.

\citet{2020ApJ...893...35D} and \citet{2020arXiv200500023K} recently developed and performed such population analyses explicitly designed to accommodate a potential BBH subpopulation within the mass gap, allowing for the possibility that these systems form via hierarchical mergers of the below-mass-gap subpopulation~\citep{2017ApJ...840L..24F,2017PhRvD..95l4046G}. As discussed in Section~\ref{sec:intro}, \citet{GW190521:implications} applied the analysis proposed by~\citet{2020arXiv200500023K}, analyzing GW190521 jointly with LIGO/Virgo's ten BBH observations from the first two observing runs, GWTC-1~\citep{2019PhRvX...9c1040A}. However, the BBH population inferred from GWTC-1 does not constrain the rate of mergers above the gap, and so these population analyses did not allow for the possibility of a first-generation BH beyond the PISN gap. \citet{GW190521:implications} noted that allowing for this possibility would not significantly affect the conclusions from the hierarchical merger analysis, which found that a first-generation origin for GW190521 is slightly preferred even without the additional probability of $m_1$ lying above the gap.

Motivated by the expectation that BBH systems involving two mass-gap BHs are a few orders of magnitude more rare than systems involving just one mass-gap BH, this work explores the assumption that the secondary mass of GW190521 is a conventional BH belonging to the component BH population that LIGO/Virgo observed in their first two observing runs.
Rather than performing a fully self-consistent population analysis that fits for both component masses of GW190521 together with previous BBH observations as in~\citet{2020ApJ...893...35D} and \citet{2020arXiv200500023K}, we pursue a simpler, less rigorous approach. In this approach, we put a population prior only on the secondary mass of GW190521, assuming it is drawn from the same population as the component masses of the BBH events observed in O1 and O2, coupled with a flat prior on $m_1$.

The mass distribution of BBHs inferred from LIGO/Virgo's first two observing runs is presented in \citet{o2pop}, with a recent update incorporating IAS detections in~\citet{2020arXiv200807014R}. In particular, the one-dimensional mass distribution describing component BHs is found to be well-fit by a power law with a variable slope, minimum mass, and maximum mass cutoff; {see for example Model B in \citet{o2pop}}. Notably, the maximum mass cutoff is well-measured at $\mmax = \mmaxModelB$, in agreement with expectations from PISN theory.

Neglecting the mass ratio distribution (although see the discussion in Section~\ref{sec:discussion}), we can use the one-dimensional mass spectrum inferred under Model B in \citet{o2pop} to construct a prior on the mass of a conventional BH (in other words, a BH that is drawn from this same population), marginalizing over the uncertainty in the population parameters. (See \citealp{2020ApJ...891L..27F} for a discussion of the sometimes subtle distinction between the component mass distribution, the primary mass distribution, and the secondary mass distribution.)
This population-informed prior is given by the posterior population distribution (PPD) inferred from~\citet{o2pop}:
\begin{equation}
\label{eq:ppd}
    p(m \mid d_\mathrm{O1+O2}) \propto \int p(m \mid \theta) p(\theta \mid d_\mathrm{O1+O2}) \mathrm{d}\theta,
\end{equation}
where $\theta = \{ \alpha_m, \beta_q, m_\mathrm{min}, \mmax \}$ are the population hyperparameters of the power-law mass distribution model (Model B) and $p(\theta \mid d_\mathrm{O1+O2})$ is the hyperposterior on these parameters inferred from the first two observing runs~\citep{o2pop}.
When we refer to a population-informed prior in the main text, we use the above PPD as a prior for $m_2$, and retain the flat prior on $m_1$.
To calculate the updated posterior on $m_1$ and $m_2$ under this prior, we reweight the posterior samples by $p(m_2 \mid d_\mathrm{O1+O2})/p_\mathrm{default}(m_1, m_2)$, where $p_\mathrm{default}(m_1, m_2)$ is the default parameter estimation prior (see Equation C1 in \citealp{o2pop}). The posteriors shown in Fig.~\ref{fig:m1-posts} are computed as weighted histograms given the original parameter estimation samples from~\citet{GW190521,GW190521:implications}, and credible intervals are calculated as weighted quantiles. Figure~\ref{fig:m1m2joint} shows a weighted kernel density estimate of the two-dimensional $m_1$, $m_2$ posterior, given the original parameter estimation samples. For plotting purposes, we extrapolate the posterior past $m_1 = 158\,\msun$ (the maximum parameter estimation sample available with the default priors) using the kernel density estimate. This extrapolation is not used for any of our numerical results. 

\section{Bayes factors between prior choices}
\label{sec:Bayesfactors}
In this section, we compare the various mass priors considered in the main text by computing their Bayes factors given the GW190521 data. We stress that it is the goal of a hierarchical Bayesian population analysis to find the common prior that best matches a collection of data (for example, a catalog of BBH events). Here, in comparing different mass priors, we perform a population analysis on only one event; see, for example, the discussion in Ref.~\citep{essick2020discriminating}. The goal of this Appendix is not to find the mass distribution that best fits the data, which would require analyzing multiple BBH events simultaneously, but to get a sense of how reasonable a given prior choice $\mathcal{H}$, $p(m_1, m_2 \mid \mathcal{H})$ is in light of the single-event likelihood, $p(d_\mathrm{GW190521} \mid m_1, m_2)$. The Bayesian evidence for model $\mathcal{H}$ given data $d$, conditioned on $d$ being detected, is~\citep{Mandel_2019,Thrane_2019,vitale2020thousand}:
\begin{equation}
\label{eq:evidence}
    p(d \mid \mathcal{H}, \mathrm{det}) = \frac{\int p(d \mid m_1, m_2) p(m_1, m_2 \mid \mathcal{H}) dm_1 dm_2}{\int P_\mathrm{det}(m_1, m_2)p(m_1, m_2 \mid \mathcal{H}) dm_1 dm_2}.
\end{equation}
The denominator of Eq.~\ref{eq:evidence} corresponds to the expected fraction of detected systems, assuming the systems are distributed according to the model $\mathcal{H}$.
In calculating this term, we follow the semi-analytic method described in~\citet{o2pop} for calculating the detection probability term $P_\mathrm{det}(m_1, m_2)$, approximating the detection threshold as a single-detector signal-to-noise ratio. This denominator varies by a factor of $\lesssim 5$ between the different prior models we consider. We calculate the numerator via importance sampling:
\begin{equation}
\label{eq:MC}
  \int p(d \mid m_1, m_2) p(m_1, m_2 \mid \mathcal{H}) dm_1 dm_2 = \left\langle \frac{p(m_1, m_2 \mid \mathcal{H})}{p_\mathrm{default}(m_1, m_2)} \right\rangle_{\{m_1, m_2\}}, 
\end{equation}
where $\langle \dots \rangle_{\{m_1, m_2\}}$ denotes an average over parameter estimation samples and $p_\mathrm{default}(m_1, m_2)$ is the prior used for parameter estimation. The standard deviation of the Monte Carlo integral in Eq.~\ref{eq:MC} can be estimated by:
\begin{equation}
   \sigma = \frac{1}{N} \sqrt{\sum_{i=1}^{N} \left(\frac{p(m_{1, i}, m_{2,i} \mid \mathcal{H})}{p_\mathrm{default}(m_{1, i}, m_{2, i})}\right)^2 - \frac{1}{N}\left(\sum_{i=1}^{N} \frac{p(m_{1, i}, m_{2,i} \mid \mathcal{H})}{p_\mathrm{default}(m_{1, i}, m_{2, i})}\right)^2 },
\end{equation}
where $N$ is the number of parameter estimation samples.
We verify that this uncertainty is smaller than $\sim 2\%$ for all models $\mathcal{H}$.

We consider three different prior models: an uninformative, flat prior on both masses,
\begin{equation}
    p(m_1, m_2 \mid \mathcal{A}, m_\mathrm{min}, \mmax) = \frac{2}{(\mmax - m_\mathrm{min})^2},
\end{equation}
an O1+O2 population-informed $m_2$ distribution (Eq.~\ref{eq:ppd}), coupled with a flat $m_1$ prior,
\begin{equation}
    p(m_1, m_2 \mid \mathcal{B}, \mmax) = p(m_2 \mid d_\mathrm{O1+O2})\frac{1}{\mmax - m_2},
\end{equation}
and an O1+O2 population-informed $m_2$ distribution, coupled with a flat $m_1$ prior restricted to $m_1 > 120\,\msun$,
\begin{equation}
    p(m_1, m_2 \mid \mathcal{C}, \mmax) = p(m_2 \mid d_\mathrm{O1+O2})\frac{1}{\mmax - 120\,\msun}.
\end{equation}
Model $\mathcal{C}$ corresponds to the ``straddling" scenario.
All cases restrict $m_\mathrm{min} < m_2 < m_1 < \mmax$.

We compute the evidence ratio between these prior choices for the GW190521 mass measurement. 
Comparing a flat, uninformative prior between $m_\mathrm{min} = 5\,\msun$ and $\mmax = 200\,\msun$ ($\mathcal{A}$) to the population-informed $m_2$ prior with a flat $m_1$ prior in the range $m_2 < m_1 < 200\,\msun$ ($\mathcal{B}$) gives a Bayes factor of $B_{\mathcal{A}/\mathcal{B}} = 6.8$ in favor of the uninformative prior. 
Meanwhile, the straddling prior $\mathcal{C}$ is favored compared to $\mathcal{B}$ by a factor of $\sim 2$; as discussed in the main text, imposing the population-informed prior on $m_2$ naturally pulls the $m_1$ posterior to sit above the gap. The Bayes factor comparing the uninformative prior to the straddling prior is $B_{\mathcal{A}/\mathcal{C}} = 3.4$ in favor of the uninformative prior.
These near-unity Bayes factors imply that the informed priors we consider in this work, which incorporate information from previous BBH observations, are reasonable alternatives to the flat, uninformative priors. 

As an additional check, we compute the Bayes factor between a double-mass-gap prior, which we take here to be model $\mathcal{A}$ but with $m_\mathrm{min} = 45\,\msun$ and $\mmax = 120\,\msun$, and the straddling mass prior $\mathcal{C}$ above. We find that, although the bulk of the GW190521 likelihood lies within the double-mass-gap mass range, the likelihood (conditioned on detection) favors the double-mass-gap prior by only a factor of $15$ compared to the straddling mass prior. As long as the prior odds for a double-mass-gap merger are smaller than $1/15$ (most theories predict prior odds smaller than $1/1000$), the posterior odds will favor that GW190521 is a straddling binary.

\end{document}